# Coloring translates and homothets of a convex body

Adrian Dumitrescu[*]     Minghui Jiang[†]

November 16, 2018

**Abstract**

We obtain improved upper bounds and new lower bounds on the chromatic number as a linear function of the clique number, for the intersection graphs (and their complements) of finite families of translates and homothets of a convex body in $\mathbb{R}^n$.

## 1 Introduction

Let us recall the following well-known hypergraph invariants for a family $\mathcal{F}$ of sets:

*clique number* $\omega(\mathcal{F})$ is the maximum number of pairwise intersecting sets in $\mathcal{F}$.

*packing number* $\nu(\mathcal{F})$ is the maximum number of pairwise disjoint sets in $\mathcal{F}$.

*clique-partition number* $\vartheta(\mathcal{F})$ is the minimum number of classes in a partition of $\mathcal{F}$ into subfamilies of pairwise intersecting sets.

*coloring number* $q(\mathcal{F})$ is the minimum number of classes in a partition of $\mathcal{F}$ into subfamilies of pairwise disjoint sets.

Let $G$ be the *intersection graph* of $\mathcal{F}$ such that the vertices in $G$ correspond to the sets in $\mathcal{F}$, one vertex for each set, and an edge connects two vertices in $G$ if and only if the corresponding two sets in $\mathcal{F}$ intersect. Then the four hypergraph invariants for $\mathcal{F}$ are respectively the same as the following four graph invariants for $G$:

*clique number* $\omega(G)$ is the maximum number of pairwise adjacent vertices (i.e., the maximum size of a clique) in $G$.

*independence number* (or *stability number*) $\alpha(G)$ is the maximum number of pairwise non-adjacent vertices (i.e., the maximum size of an independent set) in $G$.

*clique-partition number* $\vartheta(G)$ is the minimum number of classes in a partition of the vertices of $G$ into subsets of pairwise adjacent vertices.

[*]Department of Computer Science, University of Wisconsin–Milwaukee, WI 53201-0784, USA. Email: dumitres@uwm.edu. Supported in part by NSF CAREER grant CCF-0444188.
[†]Department of Computer Science, Utah State University, Logan, UT 84322-4205, USA. Email: mjiang@cc.usu.edu. Supported in part by NSF grant DBI-0743670.

*chromatic number* $\chi(G)$ is the minimum number of classes in a partition of the vertices of $G$ into subsets of pairwise non-adjacent vertices.

Let $\overline{G}$ be the *complement graph* of $G$ with the same vertices as $G$ such that two vertices are adjacent in $\overline{G}$ if and only if they are not adjacent in $G$. Then $\alpha(G) = \omega(\overline{G})$ and $\vartheta(G) = \chi(\overline{G})$.

For any family $\mathcal{F}$ of sets, we always have the following two obvious inequalities

$$\omega(\mathcal{F}) \leq q(\mathcal{F}), \quad \nu(\mathcal{F}) \leq \vartheta(\mathcal{F}). \tag{1}$$

In graph invariants, the two inequalities become

$$\omega(G) \leq \chi(G), \quad \omega(\overline{G}) \leq \chi(\overline{G}).$$

Inequalities in the opposite directions, if any, are less obvious. That is, we have only limited knowledge about possible upper bounds on the chromatic number as a function of the clique number for various classes of graphs. In this paper, we focus on finite families $\mathcal{F}$ of translates or homothets of a convex body in $\mathbb{R}^n$, and study upper bounds on the chromatic number in terms of the clique number in the intersection graphs of such families $\mathcal{F}$ and in the complement graphs. Recall that a *convex body* is a compact convex set with non-empty interior. Many similar bounds have been studied for various geometric intersection graphs and their complements since the pioneering work of Asplund and Grünbaum [3], Gyárfás [10], and Gyárfás and Lehel [11]. We refer to Kostochka [15] for a more recent survey.

**Definitions.** For two convex bodies $A$ and $B$ in $\mathbb{R}^n$, denote by $A + B = \{a + b \mid a \in A, b \in B\}$ the Minkowski sum of $A$ and $B$. For a convex body $C$ in $\mathbb{R}^n$, denote by $\lambda C = \{\lambda c \mid c \in C\}$ the *scaled copy* of $C$ by a factor of $\lambda \in \mathbb{R}$, denote by $C + p = \{c + p \mid c \in C\}$ the *translate* of $C$ by a vector from the origin to a point $p \in \mathbb{R}^n$, and denote by $\lambda C + p = \{\lambda c + p \mid c \in C\}$ the *homothet* of $C$ obtained by first scaling $C$ by a factor of $\lambda$ then translating the scaled copy by a vector from the origin to $p$. Also denote by $-C = \{-c \mid c \in C\}$ the *reflexion* of $C$ about the origin, and write $C - C$ for $C + (-C)$.

We review some standard definitions concerning packing densities; see [4, Section 1.1]. A family $\mathcal{F}$ of convex bodies is a *packing* in a domain $Y \subseteq \mathbb{R}^n$ if $\bigcup_{C \in \mathcal{F}} C \subseteq Y$ and the convex bodies in $\mathcal{F}$ are pairwise interior-disjoint. Denote by $\mu(S)$ the Lebesgue measure of a compact set $S$ in $\mathbb{R}^n$, i.e., area in the plane, or volume in the space. Define the *density* of a packing $\mathcal{F}$ relative to a bounded domain $Y$ as

$$\rho(\mathcal{F}, Y) := \frac{\sum_{C \in \mathcal{F}} \mu(C \cap Y)}{\mu(Y)}. \tag{2}$$

When $Y = \mathbb{R}^n$ is the whole space, define the *upper density* of $\mathcal{F}$ as

$$\overline{\rho}(\mathcal{F}, \mathbb{R}^n) := \limsup_{r \to \infty} \rho(\mathcal{F}, B^n(r)),$$

where $B^n(r)$ denote a ball of radius $r$ centered at the origin (since we are taking the limit as $r \to \infty$, a hypercube of side length $r$ can be used instead of a ball of radius $r$). For a convex body $C$ in $\mathbb{R}^n$, define the *packing density* of $C$ as

$$\delta(C) := \sup_{\mathcal{F} \text{ packing}} \overline{\rho}(\mathcal{F}, \mathbb{R}^n),$$

where $\mathcal{F}$ ranges over all packings in $\mathbb{R}^n$ with congruent copies of $C$. If the members of $\mathcal{F}$ are restricted to translates of $C$, then we have the *translative packing density* $\delta_T(C)$, which is invariant under any non-singular affine transformation of $C$.

**Translates and homothets of a convex body.** For $n = 1$, a convex body in $\mathbb{R}^n$ is an interval, and the intersection graph of a finite family $\mathcal{F}$ of translates or homothets of an interval is an interval graph. Since interval graphs and their complements are perfect graphs [9], we always have perfect equalities $\omega(\mathcal{F}) = q(\mathcal{F})$ and $\nu(\mathcal{F}) = \vartheta(\mathcal{F})$.

Henceforth let $n \geq 2$. Let $\mathcal{T}$ be a finite family of translates of a convex body in $\mathbb{R}^n$. Let $\mathcal{H}$ be a finite family of homothets of a convex body in $\mathbb{R}^n$. Kostochka [15] proved that

1. if $\omega(\mathcal{T}) = k$, then $q(\mathcal{T}) \leq n(2n)^{n-1}(k-1) + 1$, and
2. if $\omega(\mathcal{H}) = k$, then $q(\mathcal{H}) \leq (2n)^n(k-1) + 1$.

Kim and Nakprasit [14] proved the complementary results[1] that

1. if $\nu(\mathcal{T}) = k$, then $\vartheta(\mathcal{T}) \leq n(2n)^{n-1}(k-1) + 1$, and
2. if $\nu(\mathcal{H}) = k$, then $\vartheta(\mathcal{H}) \leq (2n)^n(k-1) + 1$.

For the planar case $n = 2$, there exist better bounds $q(\mathcal{T}) \leq 3\omega(\mathcal{T}) - 2$ and $q(\mathcal{H}) \leq 6\omega(\mathcal{T}) - 6$ by Kim, Kostochka, and Nakprasit [13], and $\vartheta(\mathcal{T}) \leq 3\nu(\mathcal{T}) - 2$ and $\vartheta(\mathcal{H}) \leq 6\nu(\mathcal{H}) - 5$ by Kim and Nakprasit [14].

For translates, we obtain the following improved bounds:

**Theorem 1.** *Let $\mathcal{T}$ be a finite family of translates of a convex body in $\mathbb{R}^n$, $n \geq 2$. Let $t_n = (n+1)^{n-1} \lceil \frac{n+1}{2} \rceil$. Then $q(\mathcal{T}) \leq t_n \omega(\mathcal{T})$ and $\vartheta(\mathcal{T}) \leq t_n \nu(\mathcal{T})$.*

Note that for all $n \geq 2$, the multiplicative factors $t_n = (n+1)^{n-1} \lceil \frac{n+1}{2} \rceil$ in Theorem 1 are exponentially smaller than the corresponding factors $n(2n)^{n-1}$ in the previous bounds [15, 14].

For two convex bodies $A$ and $B$ in $\mathbb{R}^n$, denote by $\kappa(A, B)$ the smallest number $\kappa$ such that $A$ can be covered by $\kappa$ translates of $B$. For homothets, we obtain the following bounds:

**Theorem 2.** *Let $\mathcal{H}$ be a finite family of homothets of a convex body $C$ in $\mathbb{R}^n$, $n \geq 2$. Let $h(C) = \kappa(C - C, C)$. Then $q(\mathcal{H}) \leq h(C)(\omega(\mathcal{H}) - 1) + 1$ and $\vartheta(\mathcal{H}) \leq h(C)(\nu(\mathcal{H}) - 1) + 1$.*

It remains to bound $\kappa(C - C, C)$. For a convex body $C$ in $\mathbb{R}^n$, denote by $\theta_T(C)$ the infimum of the covering density of $\mathbb{R}^n$ by translates of $C$. According to a result of Rogers [17], $\theta_T(C) < n \ln n + n \ln \ln n + 5n = O(n \log n)$ for any convex body $C$ in $\mathbb{R}^n$. The following lemma collects the previously known upper bounds on $\kappa(C - C, C)$ from [7]:

**Lemma 1** (Danzer and Rogers, 1963). *Let $C$ be a convex body $C$ in $\mathbb{R}^n$, $n \geq 2$. Then $\kappa(C-C,C) \leq 3^{n+1}2^n(n+1)^{-1}\theta_T(C) = O(6^n \log n)$. Moreover, if $C$ is centrally symmetric, then $\kappa(C - C, C) = \kappa(2C, C) \leq \min\{5^n, 3^n \theta_T(C)\} = O(3^n n \log n)$.*

Note that by Lemma 1, the multiplicative factors $h(C) = O(6^n \log n)$ in Theorem 2 are exponentially smaller than the corresponding factors $(2n)^n$ in the previous bounds [15, 14].

For the coloring problem on finite families $\mathcal{T}$ of translates of a convex body $C$ in $\mathbb{R}^n$, Kostochka [15] noted that, by the following old result of Minkowski, we can assume that $C$ is centrally symmetric:

**Lemma 2** (Minkowski, 1902). *Let $a$ and $b$ be two points and let $C$ be a convex body in $\mathbb{R}^n$, $n \geq 2$. Then $(C + a) \cap (C + b) \neq \emptyset$ if and only if $(\frac{1}{2}(C - C) + a) \cap (\frac{1}{2}(C - C) + b) \neq \emptyset$.*

---
[1]Kim and Nakprasit [14] stated their result as $\vartheta(\mathcal{T}) \leq \lceil n_- \rceil \lceil 2n_- \rceil^{n-1}(k-1) + 1$ and $\vartheta(\mathcal{H}) \leq \lceil 2n_- \rceil^n(k-1) + 1$, where $n_- = (n^2 - n + 1)^{1/2}$. But since $n - 1/2 < n_- \leq n$ for all $n \geq 1$, we indeed have $\lceil n_- \rceil = n$ and $\lceil 2n_- \rceil = 2n$.

Note that if $C$ is a convex body, then $\frac{1}{2}(C - C)$ is a centrally symmetric convex body, and $\frac{1}{2}(C-C) - \frac{1}{2}(C-C) = C - C$. Thus, by Theorem 2 and Lemma 2, we have the following corollary:

**Corollary 1.** *Let $\mathcal{T}$ be a finite family of translates of a convex body $C$ in $\mathbb{R}^n$, $n \geq 2$. Let $t(C) = \kappa(C - C, \frac{1}{2}(C - C))$. Then $q(\mathcal{T}) \leq t(C)(\omega(\mathcal{T}) - 1) + 1$ and $\vartheta(\mathcal{T}) \leq t(C)(\nu(\mathcal{T}) - 1) + 1$.*

Note that by Lemma 1, we have $t(C) = O(3^n n \log n)$. Thus, for sufficiently large $n$, the upper bounds in Corollary 1 are better than those in Theorem 1.

For a convex body $C$ in $\mathbb{R}^n$, define

$$r_T(C) = \sup_{\mathcal{T}} \frac{q(\mathcal{T})}{\omega(\mathcal{T})}, \quad \overline{r}_T(C) = \sup_{\mathcal{T}} \frac{\vartheta(\mathcal{T})}{\nu(\mathcal{T})}, \quad r_H(C) = \sup_{\mathcal{H}} \frac{q(\mathcal{H})}{\omega(\mathcal{H})}, \quad \overline{r}_H(C) = \sup_{\mathcal{H}} \frac{\vartheta(\mathcal{H})}{\nu(\mathcal{H})},$$

where $\mathcal{T}$ ranges over all finite families of translates of $C$, and $\mathcal{H}$ ranges over all finite families of homothets of $C$. Clearly, $r_T(C) \leq r_H(C)$ and $\overline{r}_T(C) \leq \overline{r}_H(C)$. Our results in Theorem 1, Theorem 2, and Corollary 1 can be summarized as follows:

$$r_T(C), \overline{r}_T(C) \leq \min\left\{(n+1)^{n-1}\left\lceil \tfrac{n+1}{2} \right\rceil,\ 5^n,\ 3^n \theta_T\left(\tfrac{1}{2}(C-C)\right)\right\} \tag{3}$$

$$r_H(C), \overline{r}_H(C) \leq 3^{n+1} 2^n (n+1)^{-1} \theta_T(C) \tag{4}$$

A natural question is whether the four ratios $r_T(C)$, $\overline{r}_T(C)$, $r_H(C)$, and $\overline{r}_H(C)$ need to be exponential in $n$. The following theorem gives a positive answer:

**Theorem 3.** *Let $C$ be a convex body in $\mathbb{R}^n$, $n \geq 2$. Then $r_H(C) \geq r_T(C) \geq 1/\delta_T(C)$ and $\overline{r}_H(C) \geq \overline{r}_T(C) \geq 1/\delta_T(C)$, where $\delta_T(C)$ is the translative packing density of $C$. In particular, if $C$ is the unit ball $B^n$ in $\mathbb{R}^n$, then $r_H(C) \geq r_T(C) \geq 2^{(0.599 \pm o(1))n}$ and $\overline{r}_H(C) \geq \overline{r}_T(C) \geq 2^{(0.599 \pm o(1))n}$ as $n \to \infty$.*

Note that our Theorem 3 gives the first general lower bounds for any convex body $C$ in $\mathbb{R}^n$, $n \geq 2$. Moreover, it gives the first lower bounds on these ratios that are exponential in the dimension $n$. Only a constant lower bound on $r_T(C)$ was previously known for the special case that $C$ is an axis-parallel square [15, 1]. We discuss this case next.

**Axis-parallel unit squares.** An interesting special case of the coloring problem is for finite families $\mathcal{F}$ of axis-parallel unit squares in the plane. Akiyama, Hosono, and Urabe [2] proved that if $\omega(\mathcal{F}) = 2$, then $q(\mathcal{F}) \leq 3$, and conjectured that, in general, if $\omega(\mathcal{F}) = k$, then $q(\mathcal{F}) \leq k+1$. Ahlswede and Karapetyan [1] recently gave a construction that disproves this conjecture. Their construction consists of a family $\mathcal{F}_k$ of squares for each $k \geq 1$, which corresponds to an intersection graph that can be obtained by "replacing each vertex of a pentagon ($C_5$) by a $k$-clique". Ahlswede and Karapetyan claimed that the family $\mathcal{F}_k$ satisfies $q(\mathcal{F}_k) = 3k$ and $\omega(\mathcal{F}_k) = 2k$, and hence gives a lower bound of $3/2$ on the multiplicative factor in the linear upper bound. On the other hand, Kostochka [15, p. 132] mentioned a lower bound of only $5/4$ (for translates of any convex body in the plane), but gave no details and no references. The following theorem resolves this discrepancy by showing that the family $\mathcal{F}_k$ in the construction by Ahlswede and Karapetyan indeed disproves the conjecture of Akiyama, Hosono, and Urabe, although it only satisfies $q(\mathcal{F}_k) = \lceil \frac{5}{2}k \rceil$ and $\omega(\mathcal{F}_k) = 2k$:

**Theorem 4.** *For every positive integer $k$, there is a family $\mathcal{F}_k$ of axis-parallel unit squares in the plane such that $\omega(\mathcal{F}_k) = 2k$ and $q(\mathcal{F}_k) = \lceil \frac{5}{2}k \rceil$, and there is a family $\mathcal{F}'_k$ of axis-parallel unit squares in the plane such that $\nu(\mathcal{F}'_k) = 2k$ and $\vartheta(\mathcal{F}'_k) = 3k$.*

For any finite family $\mathcal{F}$ of axis-parallel unit hypercubes in $\mathbb{R}^n$, Perepelitsa [16] showed that if $\omega(\mathcal{F}) = k$, then $q(\mathcal{F}) \leq 2^{n-1}(k-1) + 1$. Since $\kappa(C - C, C) = \kappa(2C, C) = 2^n$ for a hypercube $C$ in $\mathbb{R}^n$, Theorem 2 implies that if $\nu(\mathcal{F}) = k$, then $\vartheta(\mathcal{F}) \leq 2^{n-1}(k-1) + 1$ too. In particular, for any finite family $\mathcal{F}$ of axis-parallel unit squares in the plane, we have $q(\mathcal{F}) \leq 2\omega(\mathcal{F}) - 1$ and $\vartheta(\mathcal{F}) \leq 2\nu(\mathcal{F}) - 1$. By Theorem 4, the multiplicative factors of 2 in these two inequalities cannot be improved to below $\frac{5}{4}$ and $\frac{3}{2}$, respectively. It is interesting that the current best lower bounds for the two factors are different.

## 2  Upper bounds for translates of a convex body in $\mathbb{R}^n$

In this section we prove Theorem 1. Let $\mathcal{T}$ be a finite family of translates of a convex body $C$ in $\mathbb{R}^n$, $n \geq 2$. Let $P$ and $Q$ be two homothetic parallelepipeds with ratio $n$ such that $P \subseteq C \subseteq Q$, as guaranteed by the following result of Chakerian and Stein [6]:

**Lemma 3** (Chakerian and Stein, 1967). *Let $C$ be a convex body in $\mathbb{R}^n$. Then $C$ contains a parallelepiped $P$ such that some translate of $nP$ contains $C$.*

Since the intersection graph of $\mathcal{T}$ is invariant under any affine transformation of $\mathbb{R}^n$, we can assume without loss of generality that $P$ is an axis-parallel unit hypercube centered at the origin, and that $Q$ is an axis-parallel hypercube of side length $n$. Then each $C$-translate $C_p = C + p$ in $\mathcal{T}$ is specified by a *reference point* $p$ that is the center of the corresponding $P$-translate. We first consider a special case of the coloring problem in the following lemma:

**Lemma 4.** *Let $\mathcal{T}_\ell$ be a subfamily of $C$-translates in $\mathcal{T}$ whose corresponding $P$-translates intersect a common line $\ell$ parallel to the axis $x_n$. Let $c_n = \lceil \frac{n+1}{2} \rceil$. Then $q(\mathcal{T}_\ell) \leq c_n \omega(\mathcal{T}_\ell)$ and $\vartheta(\mathcal{T}_\ell) \leq c_n \nu(\mathcal{T}_\ell)$.*

*Proof.* For each integer $j$, let $U_j$ be the axis-parallel unit cube whose center is on the line $\ell$ and has $x_n$-coordinate $j$. Note that the reference point of each $C$-translate in $\mathcal{T}_\ell$ is contained in some unit cube $U_j$. Let $\mathcal{T}_c$ be the subfamily of $C$-translates in $\mathcal{T}_\ell$ whose reference points are in the unit cubes $U_j$ with $j \bmod c_n = c$. We will show that the complement of the intersection graph of each subfamily $\mathcal{T}_c$, $0 \leq c \leq c_n - 1$, is a comparability graph.

Define a relation $\prec$ on the $C$-translates in $\mathcal{T}_c$ such that $C_1 \prec C_2$ if and only if (i) $C_1$ and $C_2$ are disjoint, and (ii) the reference point of $C_1$ has a smaller $x_n$-coordinate than the reference point of $C_2$. Then the complement of the intersection graph of $\mathcal{T}_c$ has an edge between two vertices $C_1$ and $C_2$ if and only if either $C_1 \prec C_2$ or $C_2 \prec C_1$. It is clear that the relation $\prec$ is irreflexive and asymmetric. We next show that $\prec$ is also transitive, and is thus a strict partial order.

Let $C_1, C_2, C_3$ be any three $C$-translates in $\mathcal{T}_c$ such that $C_1 \prec C_2$ and $C_2 \prec C_3$. Refer to Figure 1 for an example in the plane. We will show that $C_1 \prec C_3$. Let $U_{j_1}, U_{j_2}, U_{j_3}$ be three unit cubes containing the reference points of $C_1, C_2, C_3$, respectively. Since any two $C$-translates with reference points in the same unit cube $U_j$ must intersect each other, the condition $C_1 \prec C_2$ implies that $j_1 < j_2$. Moreover we must have $j_1 \leq j_2 - c_n$ since $j_1 \equiv j_2 \pmod{c_n}$. Similarly, the condition $C_2 \prec C_3$ implies that $j_2 \leq j_3 - c_n$. It follows that $j_3 - j_1 \geq 2c_n \geq n+1$. The distance between the references points of $C_1$ and $C_3$ is at least the distance between the centers of $U_{j_1}$ and $U_{j_3}$ minus 1, which is at least $n$. This implies that $C_1$ and $C_3$ are disjoint, since each $C$-translate is contained in an axis-parallel hypercube of side length $n$. Thus $C_1 \prec C_3$ because (i) $C_1$ and $C_3$ are disjoint, and (ii) the reference point of $C_1$ has smaller $x_n$-coordinate than the reference point of $C_3$. We have shown that $\prec$ is a strict partial order. Consequently, the complement of the intersection graph of each subfamily $\mathcal{T}_c$, $0 \leq c \leq c_n - 1$, is a comparability graph.

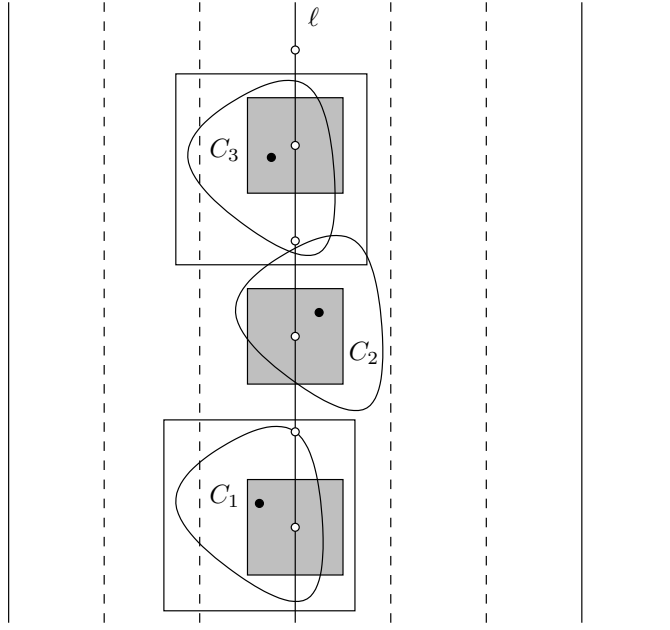

Figure 1: The subfamily of $C$-translates whose corresponding $P$-translates intersecting a common line $\ell$ parallel to the $y$-axis. The centers of the unit cubes $U_j$ are the white dots; the three unit cubes $U_{j_1}, U_{j_2}, U_{j_3}$ are shaded. The reference points of the three $C$-translates $C_1, C_2, C_3$ are the three black dots in the three unit cubes $U_{j_1}, U_{j_2}, U_{j_3}$, respectively. Each $C$-translate is contained in an axis-parallel square of side length 2 centered at its reference point. The vertical lines are equally spaced, with a distance of 3 between consecutive solid lines.

It is well-known that comparability graphs and their complements are perfect graphs [9]. So we have $q(\mathcal{T}_c) = \omega(\mathcal{T}_c)$ and $\vartheta(\mathcal{T}_c) = \nu(\mathcal{T}_c)$ for all $0 \leq c \leq c_n - 1$. Therefore,

$$q(\mathcal{T}_\ell) \leq \sum_c q(\mathcal{T}_c) = \sum_c \omega(\mathcal{T}_c) \leq \sum_c \omega(\mathcal{T}_\ell) = c_n \, \omega(\mathcal{T}_\ell).$$
$$\vartheta(\mathcal{T}_\ell) \leq \sum_c \vartheta(\mathcal{T}_c) = \sum_c \nu(\mathcal{T}_c) \leq \sum_c \nu(\mathcal{T}_\ell) = c_n \, \nu(\mathcal{T}_\ell). \qquad \Box$$

For each point $(a_1, \ldots, a_{n-1}) \in \mathbb{R}^{n-1}$, denote by $\langle a_1, \ldots, a_{n-1} \rangle$ the following line in $\mathbb{R}^n$ that is parallel to the axis $x_n$:

$$\{ (x_1, \ldots, x_n) \mid (x_1, \ldots, x_{n-1}) = (a_1, \ldots, a_{n-1}) \}.$$

Now consider the following (infinite) set $\mathcal{L}$ of (periodical) parallel lines:

$$\mathcal{L} = \{ \langle j_1 + b_1, \ldots, j_{n-1} + b_{n-1} \rangle \mid (j_1, \ldots, j_{n-1}) \in \mathbb{Z}^{n-1} \},$$

where the offset $(b_1, \ldots, b_{n-1}) \in \mathbb{R}^{n-1}$ is chosen such that no line in $\mathcal{L}$ is tangent to the $P$-translate of any $C$-translate in $\mathcal{T}$. Recall that $P$ and $Q$ are axis-parallel hypercubes of side lengths 1 and $n$, respectively. Thus we have the following two properties:

1. For any $C$-translate in $\mathcal{T}$, the corresponding $P$-translate intersects exactly one line in $\mathcal{L}$.

2. For any two $C$-translates in $\mathcal{T}$, if the two corresponding $P$-translates intersect two different lines in $\mathcal{L}$ at distance at least $n+1$ along some axis $x_i$, $1 \leq i \leq n-1$, then the two $C$-translates are disjoint.

Partition $\mathcal{T}$ into subfamilies $\mathcal{T}[j_1, \ldots, j_{n-1}]$ of $C$-translates whose corresponding $P$-translates intersect a common line $\langle j_1 + b_1, \ldots, j_{n-1} + b_{n-1} \rangle$. By Lemma 4, the coloring number and the clique-partition number of each subfamily $\mathcal{T}[j_1, \ldots, j_{n-1}]$ are at most $c_n$ times its clique number and its packing number, respectively. For each $(k_1, \ldots, k_{n-1}) \in \{0, 1, \ldots, n\}^{n-1}$, let $\mathcal{T}_\cup[k_1, \ldots, k_{n-1}]$ be the union of the (pairwise-disjoint) subfamilies $\mathcal{T}[j_1, \ldots, j_{n-1}]$ with $j_i \equiv k_i \pmod{n+1}$ for all $1 \leq i \leq n-1$. Again refer to Figure 1 for an example in the plane. Then,

$$\begin{aligned} q(\mathcal{T}_\cup[k_1, \ldots, k_{n-1}]) &= \max_{j_i \equiv k_i} q(\mathcal{T}[j_1, \ldots, j_{n-1}]) \\ &\leq \max_{j_i \equiv k_i} c_n \, \omega(\mathcal{T}[j_1, \ldots, j_{n-1}]) \\ &= c_n \, \omega(\mathcal{T}_\cup[k_1, \ldots, k_{n-1}]) \\ &\leq c_n \, \omega(\mathcal{T}) \end{aligned}$$

and

$$\begin{aligned} \vartheta(\mathcal{T}_\cup[k_1, \ldots, k_{n-1}]) &= \sum_{j_i \equiv k_i} \vartheta(\mathcal{T}[j_1, \ldots, j_{n-1}]) \\ &\leq \sum_{j_i \equiv k_i} c_n \, \nu(\mathcal{T}[j_1, \ldots, j_{n-1}]) \\ &= c_n \, \nu(\mathcal{T}_\cup[k_1, \ldots, k_{n-1}]) \\ &\leq c_n \, \nu(\mathcal{T}). \end{aligned}$$

Consequently,

$$q(\mathcal{T}) \leq \sum_{k_1, \ldots, k_{n-1}} q(\mathcal{T}_\cup[k_1, \ldots, k_{n-1}]) \leq \sum c_n \, \omega(\mathcal{T}) = (n+1)^{n-1} c_n \, \omega(\mathcal{T}) = t_n \, \omega(\mathcal{T}),$$

$$\vartheta(\mathcal{T}) \leq \sum_{k_1, \ldots, k_{n-1}} \vartheta(\mathcal{T}_\cup[k_1, \ldots, k_{n-1}]) \leq \sum c_n \, \nu(\mathcal{T}) = (n+1)^{n-1} c_n \, \nu(\mathcal{T}) = t_n \, \nu(\mathcal{T}).$$

This completes the proof of Theorem 1.

## 3 Upper bounds for homothets of a convex body in $\mathbb{R}^n$

In this section we prove Theorem 2. Let us define one more hypergraph invariant for a family $\mathcal{F}$ of sets:

*transversal number* $\tau(\mathcal{F})$ is the minimum cardinality of a set of elements that intersects all sets in $\mathcal{F}$.

Since any subfamily of $\mathcal{F}$ that share a common element corresponds to a clique the intersection graph of $\mathcal{F}$, we have the following inequality in addition to (1):

$$\vartheta(\mathcal{F}) \leq \tau(\mathcal{F}). \tag{5}$$

For the special case that $\mathcal{F}$ is a family of axis-parallel boxes in $\mathbb{R}^n$, we indeed have $\vartheta(\mathcal{F}) = \tau(\mathcal{F})$ since any subfamily of pairwise-intersecting axis-parallel boxes must share a common point. We will use the following lemma from a related work of ours on transversal numbers [8]:

**Lemma 5** (Dumitrescu and Jiang, 2009). *Let $\mathcal{H}$ be a finite family of homothets of a convex body $C$ in $\mathbb{R}^n$, $n \geq 2$. Let $C_1$ be the smallest homothet in $\mathcal{H}$, and let $\mathcal{H}_1$ be the subfamily of homothets in $\mathcal{H}$ that intersect $C_1$ ($\mathcal{H}_1$ includes $C_1$ itself). Then $\tau(\mathcal{H}_1) \leq \kappa(C - C, C)$.*

By inequality (5), we immediately have the following corollary:

**Corollary 2.** *Let $\mathcal{H}$ be a finite family of homothets of a convex body $C$ in $\mathbb{R}^n$, $n \geq 2$. Let $C_1$ be the smallest homothet in $\mathcal{H}$, and let $\mathcal{H}_1$ be the subfamily of homothets in $\mathcal{H}$ that intersect $C_1$ ($\mathcal{H}_1$ includes $C_1$ itself). Then $\vartheta(\mathcal{H}_1) \leq \kappa(C - C, C)$.*

We first bound $q(\mathcal{H})$ in terms of $\omega(\mathcal{H})$. As in Corollary 2, let $C_1$ be the smallest homothet in $\mathcal{H}$, and let $\mathcal{H}_1$ be the subfamily of homothets in $\mathcal{H}$ that intersect $C_1$. Consider any partition of $\mathcal{H}_1$ into at most $\vartheta(\mathcal{H}_1)$ classes of pairwise-intersecting homothets. Add $C_1$ to each class if it is not already there. Then in each class the homothets are pairwise-intersecting, and the number of homothets except $C_1$ is at most $\omega(\mathcal{H}_1) - 1$. Thus $C_1$ intersects a total of at most $\vartheta(\mathcal{H}_1)(\omega(\mathcal{H}_1) - 1) \leq \kappa(C - C, C)(\omega(\mathcal{H}) - 1)$ other homothets in $\mathcal{H}$. By a standard recursive argument, it follows that

$$q(\mathcal{H}) \leq \kappa(C - C, C)(\omega(\mathcal{H}) - 1) + 1.$$

We next bound $\vartheta(\mathcal{H})$ in terms of $\nu(\mathcal{H})$. Consider the following greedy partition of $\mathcal{H}$: first find in $\mathcal{H}$ the smallest homothet $C_1$ and the subfamily $\mathcal{H}_1$ of homothets that intersect $C_1$, next find in $\mathcal{H} \setminus \mathcal{H}_1$ the smallest homothet $C_2$ and the subfamily $\mathcal{H}_2$ of homothets that intersect $C_2$, and so on. Let $\mathcal{H} = \mathcal{H}_1 \cup \cdots \cup \mathcal{H}_k$ be the resulting partition. Then $k \leq \nu(\mathcal{H})$ since the homothets $C_i$ are pairwise-disjoint. By Corollary 2, $\vartheta(\mathcal{H}_i) \leq \kappa(C - C, C)$ for each $\mathcal{H}_i$ in the partition. Moreover, if $k = \nu(\mathcal{H})$, then we must have $\vartheta(\mathcal{H}_k) = 1$ since otherwise there would be more than $k$ pairwise-disjoint homothets in $\mathcal{H}$. Thus

$$\vartheta(\mathcal{H}) \leq \sum_{i=1}^{k} \vartheta(\mathcal{H}_i) \leq \left( \sum_{i=1}^{\nu(\mathcal{H})-1} \kappa(C - C, C) \right) + 1 = \kappa(C - C, C)(\nu(\mathcal{H}) - 1) + 1.$$

This completes the proof of Theorem 2.

## 4 Lower bounds for translates of a convex body in $\mathbb{R}^n$

In this section we prove Theorem 3. Let $C$ be a convex body in $\mathbb{R}^n$ and $m$ be a positive integer. We will show that $r_T(C) \geq 1/\delta_T(C)$ and $\overline{r}_T(C) \geq 1/\delta_T(C)$ by constructing a finite family $\mathcal{F}_m$ of $m^{2n}$ translates of $C$, such that

$$\lim_{m \to \infty} \frac{q(\mathcal{F}_m)}{\omega(\mathcal{F}_m)} \geq \frac{1}{\delta_T(C)}, \tag{6}$$

and

$$\lim_{m \to \infty} \frac{\vartheta(\mathcal{F}_m)}{\nu(\mathcal{F}_m)} \geq \frac{1}{\delta_T(C)}. \tag{7}$$

By Lemma 2, we can assume that $C$ is centrally symmetric and is centered at the origin. We will use the following isodiametric inequality due to Busemann [5, p. 241, (2.2)]:

**Lemma 6** (Busemann, 1947). *Let $C$ be a centrally symmetric convex body in $\mathbb{R}^n$. Let $\mathbb{M}^n$ be the Minkowski space in which $C$ is a ball of unit radius. For any measurable set $S$ in $\mathbb{R}^n$ of Minkowski diameter at most $2$ in $\mathbb{M}^n$, the Lebesgue measure of $S$ in $\mathbb{R}^n$ is at most the Lebesgue measure of $C$ in $\mathbb{R}^n$.*

Let $\mathcal{F}_m$ be a family of translates of $C$

$$\mathcal{F}_m := \{C + t \mid t \in T_m\}$$

corresponding to a set $T_m$ of $m^{2n}$ regularly placed reference points

$$T_m := \{(t_1/m, \ldots, t_n/m) \mid (t_1, \ldots, t_n) \in \mathbb{Z}^n, 1 \leq t_1, \ldots, t_n \leq m^2\}.$$

Let $U_m$ be an axis-parallel hypercube of side length $1/m$ that is centered at the origin. Observe that $U_m + T_m$ is an axis-parallel hypercube of side length $m$.

We first obtain a lower bound on $\vartheta(\mathcal{F}_m)$. Note that any two translates of $C$ in $\mathbb{R}^n$ intersect if and only if the Minkowski distance between their centers is at most 2 in $\mathbb{M}^n$. Thus any subset of pairwise intersecting translates of $C$ in $\mathcal{F}_m$ corresponds to a subset of points of Minkowski diameter at most 2 in $T_m$, and reciprocally. Consider a partition of $\mathcal{F}_m$ into $\vartheta(\mathcal{F}_m)$ subsets of pairwise intersecting translates of $C$, and let $T_m^i \subseteq T_m$, $1 \leq i \leq \vartheta(\mathcal{F}_m)$, be the corresponding subsets of Minkowski diameter at most 2. Then the hypercube $U_m + T_m$ is covered by the union of the subsets $U_m + T_m^i$, $1 \leq i \leq \vartheta(\mathcal{F}_m)$. Let $S_m \subseteq T_m$ be a maximum-cardinality subset of points of Minkowski diameter at most 2. Then, by a volume argument, we have

$$\vartheta(\mathcal{F}_m) \geq \frac{\mu(U_m + T_m)}{\mu(U_m + S_m)}. \tag{8}$$

We next obtain an upper bound on $\nu(\mathcal{F}_m)$. Let $B$ be the smallest axis-parallel box containing $C$. For each point $t \in T_m$, the corresponding translate $C + t \in \mathcal{F}_m$ satisfies $C + t \subseteq C + T_m \subseteq B + T_m$. Recall our definition (2) that $\rho(\mathcal{F}, Y)$ is the density of a family $\mathcal{F}$ of convex bodies relative to a bounded domain $Y \subseteq \mathbb{R}^n$. Let $\mathcal{I}_m \subseteq \mathcal{F}_m$ be a maximum-cardinality packing in $B + T_m$. Again, by a volume argument, we have

$$\nu(\mathcal{F}_m) \leq \rho(\mathcal{I}_m, B + T_m) \cdot \frac{\mu(B + T_m)}{\mu(C)}. \tag{9}$$

From (8) and (9), it follows that

$$\frac{\vartheta(\mathcal{F}_m)}{\nu(\mathcal{F}_m)} \geq \frac{1}{\rho(\mathcal{I}_m, B + T_m)} \cdot \frac{\mu(U_m + T_m)}{\mu(B + T_m)} \cdot \frac{\mu(C)}{\mu(U_m + S_m)}. \tag{10}$$

Now, taking the limit as $m \to \infty$, we clearly have $\rho(\mathcal{I}_m, B + T_m) \to \delta_T(C)$ and $\mu(U_m + T_m)/\mu(B + T_m) \to 1$. Also, as $m \to \infty$, the Minkowski diameter of $U_m + S_m$ tends to the Minkowski diameter of $S_m$, which is at most 2. It then follows by Lemma 6 that $\lim_{m \to \infty} \mu(U_m + S_m) \leq \mu(C)$. This yields (7) as desired.

To show (6) we now obtain bounds on $q(\mathcal{F}_m)$ and $\omega(\mathcal{F}_m)$. Since $q(\mathcal{F}_m)\nu(\mathcal{F}_m) \geq |\mathcal{F}_m| = |T_m|$, it follows immediately from (9) that

$$q(\mathcal{F}_m) \geq \frac{|T_m|}{\rho(\mathcal{I}_m, B + T_m)} \cdot \frac{\mu(C)}{\mu(B + T_m)}. \tag{11}$$

Recall the definition of $S_m$ before (8). Clearly,

$$\omega(\mathcal{F}_m) = |S_m|. \tag{12}$$

From (11) and (12), it follows that

$$\frac{q(\mathcal{F}_m)}{\omega(\mathcal{F}_m)} \geq \frac{1}{\rho(\mathcal{I}_m, B + T_m)} \cdot \frac{\mu(U_m + T_m)}{\mu(B + T_m)} \cdot \frac{\mu(C)}{\mu(U_m + S_m)} \cdot \frac{\mu(U_m + S_m)}{\mu(U_m + T_m)} \cdot \frac{|T_m|}{|S_m|}. \tag{13}$$

Note that $\mu(U_m + S_m) = \mu(U_m) \cdot |S_m|$ and $\mu(U_m + T_m) = \mu(U_m) \cdot |T_m|$. Hence the two inequalities (10) and (13) have the same the right-hand side. Taking the limit as $m \to \infty$ in (13) yields (6).

We have shown that $r_T(C) \geq 1/\delta_T(C)$ and $\overline{r}_T(C) \geq 1/\delta_T(C)$ for any convex body $C$ in $\mathbb{R}^n$. For the special case that $C$ is the $n$-dimensional unit ball $B^n$ in $\mathbb{R}^n$, Kabatjanskiĭ and Levenšteĭn [12] showed that $\delta_T(B^n) = \delta(B^n) \leq 2^{-(0.599 \pm o(1))n}$ and hence $1/\delta_T(B^n) \geq 2^{(0.599 \pm o(1))n}$ as $n \to \infty$; see also [4, p. 50]. This completes the proof of Theorem 3.

## 5 Lower bounds for axis-parallel unit squares

In this section we prove Theorem 4. Refer to Figure 2(a) for the construction of the family $\mathcal{F}_k$ given by Ahlswede and Karapetyan [1], $k \geq 1$.

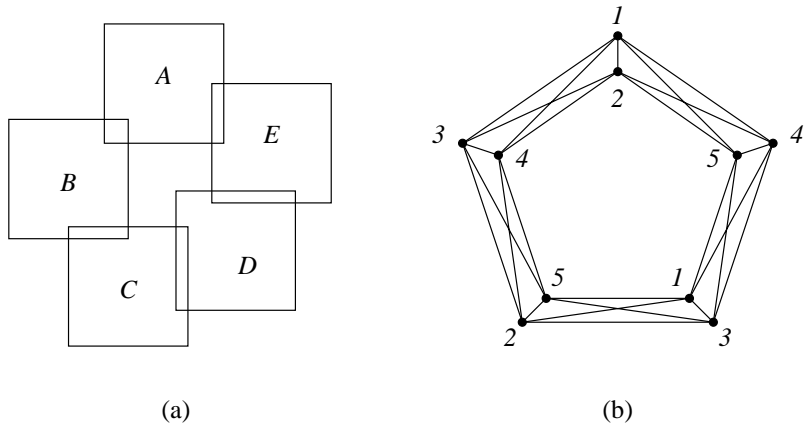

Figure 2: Lower bound construction for axis-parallel squares. (a) The family $\mathcal{F}_k$ consists of $5k$ squares, $k$ duplicates (or sufficiently close translates) of each of the five squares arranged into a 5-cycle. (b) A 5-coloring of the intersection graph of $\mathcal{F}_2$.

Let $A, B, C, D, E$ be the five groups of squares in $\mathcal{F}_k$, $k$ squares in each group. It is clear that $\omega(\mathcal{F}_k) = 2k$, which is realized by any two adjacent groups of squares, for example, $A$ and $B$. It is also clear that $q(\mathcal{F}_k) \leq 3k$. Let $Q_1, Q_2, Q_3$ be the three classes in any partition of $3k$ distinct colors, $k$ colors in each class. Then we can use $Q_1$ for $A$ and $C$, $Q_2$ for $B$ and $E$, and $Q_3$ for $D$. Ahlswede and Karapetyan [1] mistakenly assumed that $q(\mathcal{F}_k) = 3k$. We next derive the correct value of $q(\mathcal{F}_k)$.

Observe that $\nu(\mathcal{F}_k) = 2$. Thus we clearly have the lower bound $q(\mathcal{F}_k) \geq |\mathcal{F}_k|/\nu(\mathcal{F}_k) = \frac{5}{2}k$; moreover $q(\mathcal{F}_k) \geq \lceil \frac{5}{2}k \rceil$ since $q(\mathcal{F}_k)$ is an integer. To derive the matching upper bound $q(\mathcal{F}_k) \leq \lceil \frac{5}{2}k \rceil = k + k + \lceil k/2 \rceil$, we construct a coloring of $\mathcal{F}_k$ with $k$ colors from $Q_1$, $k$ colors from $Q_2$, and $\lceil k/2 \rceil$ colors from $Q_3$. Partition each color class $Q_i$, $1 \leq i \leq 3$, into two sub-classes of $Q_{i,1}$ and $Q_{i,2}$ of sizes $\lceil k/2 \rceil$ and $\lfloor k/2 \rfloor$, respectively. The coloring is as follows:

$$A: Q_{1,1} \cup Q_{1,2} \quad B: Q_{2,1} \cup Q_{2,2} \quad C: Q_{1,2} \cup Q_{3,1} \quad D: Q_{1,1} \cup Q_{2,1} \quad E: Q_{2,2} \cup Q_{3,1}$$

For coloring $D$ we use any $k$ colors from $Q_{1,1} \cup Q_{2,1}$. Observe that $D$ does not use any color in $Q_3$, and that $C$ and $E$ share the colors in $Q_{3,1}$. Refer to Figure 2(b) for the case $k = 2$.

For the second part of the theorem, let $\mathcal{F}'_k$ be $k$ *disjoint* groups of five squares each, repeating the intersection pattern in Figure 2(a). It is easy to see that $\nu(\mathcal{F}'_k) = 2k$ and $\vartheta(\mathcal{F}'_k) = 3k$. This completes the proof of Theorem 4.